\documentclass[twocolumn,showpacs,preprintnumbers,amsmath,amssymb]{revtex4}
\usepackage{lipsum}
\usepackage{tabularx}
\usepackage{array}
\usepackage{mathrsfs}
\usepackage{amssymb}
\usepackage{amsmath}
\usepackage{cases}  
\usepackage{graphicx}
\usepackage{subfigure} 
\usepackage{dcolumn}
\usepackage{bm}
\usepackage{verbatim} 
\usepackage[dvipdfm,pdfstartview=FitH,colorlinks,linkcolor=blue,anchorcolor=blue,citecolor=blue]{hyperref}
\usepackage{amsthm}  

\begin{document}

\title{Controlled-phase manipulation module for orbital-angular-momentum photon states}

\author{Fang-Xiang Wang$^{1,2,3}$}
\author{Juan Wu$^{1,2,3}$}
\author{Wei Chen$^{1,2,3}$}
\email{weich@ustc.edu.cn}
\author{Zhen-Qiang Yin$^{1,2,3}$}
\author{Shuang Wang$^{1,2,3}$}
\author{Guang-Can Guo$^{1,2,3}$}
\author{Zheng-Fu Han$^{1,2,3}$}
\email{zfhan@ustc.edu.cn}

\affiliation{$^1$CAS Key Laboratory of Quantum Information, University of Science and Technology of China, Hefei 230026, People's Republic of China\\
$^2$Synergetic Innovation Center of Quantum Information $\&$ Quantum Physics, University of Science and Technology of China, Hefei, Anhui 230026, People's Republic of China\\
$^3$State Key Laboratory of Cryptology, P. O. Box 5159, Beijing 100878, People's Republic of China}


\begin{abstract}
Phase manipulation is essential to quantum information processing, for which the orbital angular momentum (OAM) of photon is a promising high-dimensional resource. Dove prism (DP) is one of the most important element to realize the nondestructive phase manipulation of OAM photons. DP usually changes the polarization of light and thus increases the manipulation error for a spin-OAM hybrid state. DP in a Sagnac interferometer also introduces a mode-dependent global phase to the OAM mode. In this work, we implemented a high-dimensional controlled-phase manipulation module (PMM), which can compensate the mode-dependent global phase and thus preserve the phase in the spin-OAM hybrid superposition state. The PMM is stable for free running and is suitable to realize the high-dimensional controlled-phase gate for spin-OAM hybrid states. Considering the Sagnac-based structure, the PMM is also suitable for classical communication with spin-OAM hybrid light field.
\end{abstract}


\maketitle

Phase manipulation is essential to quantum information processing \cite{Kok2007,Leibfried2003,Mattle1995,Lawrence2004}. The degree of freedom of orbital angular momentum (OAM) of photons \cite{Allen1992} is an attractive resource for high-dimensional quantum information tasks, such as quantum entanglement \cite{Nagali2010,Dada2011,Fickler2012}, quantum memory \cite{Ding2016}, quantum teleportation \cite{Wang2015} and quantum communication \cite{Mirhosseini2015,Wang2016,Sit2017,Erhard2017}. The mode-dependent phase manipulation of OAM photon is commonly accomplished by utilizing a Dove prism (DP). A DP rotating with an angle $\alpha$ along its longitudinal axis will introduce an OAM-dependent phase $e^{i2l\alpha}$ to the OAM photon $|l\rangle$ \cite{Courtial1998}, where $l$ is the order of the OAM mode. Then the phase gate (Z-gate) for an OAM qudit can be implemented $|l\rangle\rightarrow e^{i2l\alpha}|l\rangle$ \cite{Zhang2016}. By utilizing this property, the Mach-Zehnder interferometer (MZI) with a DP in each arm can sort OAM photons \cite{Leach2002}. As the MZI is sensitive to the environmental disturbances, the modified Sagnac interferometers have been introduced \cite{Slussarenko2010} to overcome this shortcoming. In fact, the interferometer will also introduce an OAM-dependent global phase. For example, for a Sagnac interferometer, the output state with constructive interference is $e^{i2l\alpha}|l\rangle$. The global phase $e^{i2l\alpha}$ is insignificant for single mode OAM photons. However, for a multi-mode input OAM photon state $\sum_{l}|l\rangle$, the output state with constructive interference becomes $\sum_{l}e^{i2l\alpha}|l\rangle$. For example, for an input state $|l\rangle+|-l\rangle$, the output state becomes $i(|l\rangle-|-l\rangle)$, which is orthogonal to the input one, when $\alpha=\pi/(4l)$. Moreover, for a high-dimensional system that requires parallel manipulations in multiple paths, the final output state will becomes disordered. Additionally, the DP rotates the linear polarization into elliptical polarization and decreases the visibility of the interferometer if $\alpha\neq n\pi/2$ \cite{Padgett1999,Moreno2003}, where $n$ is an integer. For hyper-entangled states, the shortcoming may also increase the manipulation error of the states \cite{Wang2017}. The DP can be replaced by some modified prisms to overcome the polarization-dependent effect \cite{Leach2004}, but it makes the technology complex and is not commercial.

In this work, we have implemented a high-dimensional controlled-phase manipulation module (PMM) for hybrid spin-OAM photon states. The PMM compensates the mode-dependent global phase and preserves the phase of spin-OAM hybrid states. The PMM consists of a single-path Sagnac loop, two DPs and four half-wave plates (HWPs). The free running PMM is suitable for classical optical communication \cite{Willner2015} and high-dimensional quantum information processing, e.g., for hyper-entangled state manipulation in the degrees of freedom of OAM and spin.

\begin{figure}
\centering
\includegraphics[width=\linewidth]{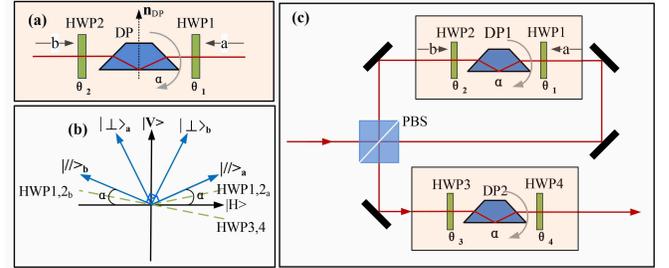}
\caption{(a) The schematic diagram of the sandwich-like structure in the Sagnac loop, in which a DP is stuck in the middle of two HWPs. (b) The rotating angles of the DP and the HWPs of (a) in the lab coordinate system. (c) The schematic setup of the phase manipulation module.}
\label{fig1}
\end{figure}

A DP rotating with an angle $\alpha$ along its longitudinal axis rotates the optical image with $2\alpha$. The interferometer (an MZI or a Sagnac one) with DPs will introduce an OAM-dependent global phase to the photon. Thus, the output state will be modulated by the  mode-dependent relative phases when the incident photon is multi-mode. Taking into account the polarization-dependent property of the DP, the linear polarization will be transformed into elliptical polarization when $\alpha$ is not $n\pi/2$ \cite{Padgett1999,Moreno2003}, which will decrease the visibility of the interferometer. The Jones matrix of a DP can be expressed as \cite{Wang2017}
\begin{equation}
J_{DP}(\alpha)=R_s(-\alpha)
\begin{pmatrix}
\sqrt{t_{//}} & 0 \\
0 & \sqrt{t_{\perp}}e^{i\Delta\varphi}
\end{pmatrix}R_s(\alpha),
\label{equ1}
\end{equation}
where $\alpha$ is the rotation angle of the DP, $t_{//(\perp)}$ is the transmission coefficient of the DP for the linear polarized light that parallel (perpendicular) to the normal of the base $\textbf{n}_{{}_{DP}}$, and $\Delta\varphi$ is the relative phase shift between the two polarization components introduced by the total internal reflection of the DP. For a spin-OAM hybrid state input into a Sagnac interferometer, the output state depends not only on the rotation angle $\alpha$ but also on the specific parameters of the DP. This shortcoming weakens the application of the DP in high-dimensional quantum information processing with spin-OAM hybrid photon. A sandwich-like structure with a DP stuck in the middle of two HWPs, as shown in Fig. \ref{fig1}(a), has been proposed to increase the visibility of the Sagnac interferometer. The sandwich-like structure works as follows in direction a \cite{Wang2017}: The rotation angles of the fast axis at its horizontal axis for both HWP1 and HWP2 are $\theta_1=\theta_2=\alpha/2$, thus, $|H\rangle|l\rangle\xrightarrow{HWP1_a}|//\rangle_a|l\rangle\xrightarrow{DP}e^{i2l\alpha}\sqrt{t_{//}}|//\rangle_a|l\rangle\xrightarrow{HWP2_a}e^{i2l\alpha}\sqrt{t_{//}}|H\rangle|l\rangle, |V\rangle|l\rangle\xrightarrow{HWP1_a}-|\perp\rangle_a|l\rangle\xrightarrow{DP}-e^{i\Delta\varphi}e^{i2l\alpha}\sqrt{t_{\perp}}|\perp\rangle_a|l\rangle\xrightarrow{HWP2_a}e^{i\Delta\varphi}e^{i2l\alpha}\sqrt{t_{\perp}}|V\rangle|l\rangle,$ where $|H\rangle (|V\rangle)$ represents the horizontal (vertical) polarization, $|//\rangle_a (|\perp\rangle_a)$ represents the polarization that is parallel (perpendicular) to the normal $\textbf{n}_{{}_{DP}}$ of the DP in direction a. The same results can be obtained for states input from direction b. What should be noted is that the axes of the DP and the HWPs are mirrored in direction b, as shown in Fig. \ref{fig1}(b). Though increasing the visibility of the BS Sagnac interferometer, the sandwich-like structure does not preserve the spin-OAM hybrid state and introduces a mode-dependent "global phase" $e^{i2l\alpha}$ \cite{Wang2017}.

Based on the sandwich-like structure above, we propose a PMM to compensate both the spin-OAM hybrid state and the "global phase". The structure of the PMM is shown in Fig. \ref{fig1}(c). The sandwich-like structure is inserted into a polarizing beam splitter (PBS) single-path Sagnac interferometer, which is cascaded by a second sandwich-like structure. The parameters of the first sandwich-like structure are set as same as that in Fig. \ref{fig1}(b). The rotation angles of DP1 and DP2 are both $\alpha_1=\alpha_2=\alpha$. The rotation angles of the fast axes of HWP3 and HWP4 are set as $\theta_3=\theta_4=-\alpha/2$. Then, for a high-dimensional spin-OAM hybrid input state $\sum_{l}(|H\rangle+|V\rangle)|l\rangle$, the PMM works as follows:
\begin{equation}
\begin{aligned}
&\sum_{l}(|H\rangle+|V\rangle)|l\rangle\\
&\xrightarrow{Sagnac}\sum_{l}e^{i2l\alpha}(\sqrt{t_{//}}|H\rangle+e^{-i4l\alpha}e^{i\Delta\varphi}\sqrt{t_{\perp}}|V\rangle)|l\rangle\\
&\xrightarrow{HWP3}\sum_{l}e^{i2l\alpha}(\sqrt{t_{//}}|\perp\rangle+e^{-i4l\alpha}e^{i\Delta\varphi}\sqrt{t_{\perp}}|//\rangle)|l\rangle\\
&\xrightarrow{DP2}\sum_{l}(\sqrt{t_{//}t_{\perp}}e^{i\Delta\varphi}|\perp\rangle+e^{-i4l\alpha}e^{i\Delta\varphi}\sqrt{t_{\perp}t_{//}}|//\rangle)|l\rangle\\
&\xrightarrow{HWP4}\sum_{l}\sqrt{t_{//}t_{\perp}}e^{i\Delta\varphi}(|H\rangle+e^{-i4l\alpha}|V\rangle)|l\rangle
\end{aligned}
\label{equ2}
\end{equation}
Hence, the mode-dependent "global phase" is compensated and the spin-OAM hybrid state becomes independent on the specific parameters of the DPs. According to Eq. \ref{equ2}, the OAM-dependent phase $e^{-i4l\alpha}$ can be controlled by the polarization of the photon, where we have ignored the OAM-independent global phase and the transmission efficiency. Thus, the PMM actually is a single-photon controlled-phase (C-phase) gate \cite{Deng2007,Wang2015}. A two dimensional C-phase gate \cite{Kok2007,Kiesel2005} applies a relative $\pi$-phase shift to the target qubit only if the controlled qubit is $|1\rangle$:
\begin{equation}
|a\rangle|b\rangle\xrightarrow{C-phase}|a\rangle(-1)^{a}|b\rangle,
\label{equ3}
\end{equation} 
where $a,b\in\{0,1\}$. For high-dimensional quantum states, a phase gate (Z-gate) is defined as \cite{Babazadeh2017}
\begin{equation}
Z=\sum_{l=0}^{D-1}|l\rangle e^{i2\pi l/N}\langle l|,
\label{equ4}
\end{equation}
where, $D$ is the dimensionality of the quantum state. The PMM proposed here introduces an OAM-dependent phase $e^{-i4l\alpha}$ to the photon only if the controlled qubit is $|V\rangle$. Thus, the PMM is a high-dimensional C-phase gate only if $\alpha=\pi/(2N)$:
\begin{equation}
CZ=\sum_{l=0}^{D-1}[(|H\rangle\langle H|)\otimes(|l\rangle\langle l|)+(|V\rangle\langle V|)\otimes(|l\rangle e^{i2\pi l/N}\langle l|)].
\label{equ5}
\end{equation}

\begin{figure*}[!hbt]
\centering
\resizebox{14.93cm}{2.8cm}{\includegraphics{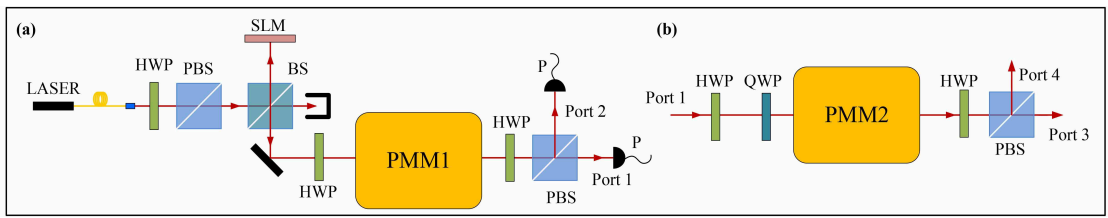}}
\caption{(a) The schematic experimental setup of the PMM. The devices after PMM1 are the detection module. P: power meter. (b) The cascading PMM (PMM2) for filtering $(4n+1)$-order OAM modes from the $(4n+3)$-order ones. It is cascaded after the last PBS of the OAM filter in (a).}
\label{fig2}
\end{figure*}

A proof-of-principle experimental setup, as shown in Fig. \ref{fig2}(a), is implemented to verify the high-dimensional C-phase gate described by Eq. \ref{equ5}. Since the first-order coherence of a single-photon can be simulated by coherent light, the TEM${}_{00}$-mode continuous wave (CW) laser with the wavelength of 780 $nm$ is used in the experiment. The spatial light modulator (SLM) and the HWP before the PMM (PMM1) are used to prepare the spin-OAM hybrid input state $\sum_{l}|(H\rangle+|V\rangle)|l\rangle$. The hybrid state is then incident into the PMM. The rotation angles of both DPs in PMM1 are set as $\pi/4$. The light is then detected by the detection module. The detection module consists of an HWP, a PBS and two power meters. The intensity profiles of the OAM modes are imaged by the CCD camera.

\begin{figure}[!htb]
\centering
\resizebox{7cm}{3.5cm}{\includegraphics{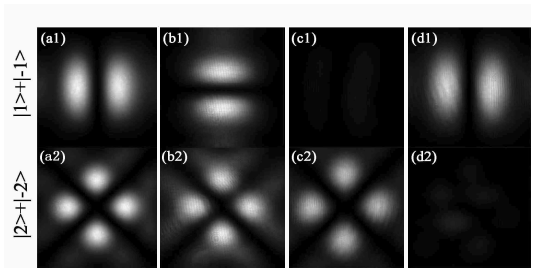}}
\caption{(a1) and (a2) are the intensity profiles of the input superposition states of $|l=1\rangle+|l=-1\rangle$ and $|l=2\rangle+|l=-2\rangle$, respectively; (b1) and (b2) are the output intensity profiles from the normal PBS single-path Sagnac interferometer; (c1) and (c2) are the intensity profiles at Port 2 output from PMM1; (d1) and (d2) are the intensity profiles at Port 1 output from PMM1.}
\label{fig3}
\end{figure}

We first compare the output states of the proposed PMM and the single-path Sagnac interferometers (without the second sandwich-like structure) \cite{Slussarenko2010,Wang2017}. Two hybrid superposition states $|\psi_1\rangle_{in}=(|H\rangle+|V\rangle)(|l=1\rangle+|l=-1\rangle)$ and $|\psi_2\rangle_{in}=(|H\rangle+|V\rangle)(|l=2\rangle+|l=-2\rangle)$ are utilized for the verification experiments. According to Refs. \cite{Slussarenko2010} and \cite{Wang2017}, the output states of the single-path Sagnac interferometer are $|\psi_1\rangle_{out}=(|H\rangle-|V\rangle)(|l=1\rangle-|l=-1\rangle$ and $|\psi_2\rangle_{out}=(|H\rangle+|V\rangle)(|l=2\rangle+|l=-2\rangle$, where the output OAM states depend on the rotation angle $\alpha$. While, from Eq. \ref{equ2}, the superposition OAM states output from the PMM are invariant. The intensity profiles of the OAM states are imaged by a CCD camera. Figure \ref{fig3} gives the experimental results. Figure \ref{fig3}(b1) shows that $|l=1\rangle+|l=-1\rangle$ has been transformed into $|l=1\rangle-|l=-1\rangle$ by the single-path Sagnac interferometer. The images of Figs. \ref{fig3}(c1), \ref{fig3}(c2), \ref{fig3}(d1) and \ref{fig3}(d2) demonstrate that the PMM preserves the OAM superposition states and controls the phase depending on the polarization. In order to evaluate the manipulation quality, we measure the sorting fidelity of the PMM, which is defined as \cite{Wang2017}
\begin{equation}
\mathcal{F}=\frac{I_{max}}{I_{max}+I_{min}},
\label{equ6}
\end{equation}
where $I_{max(min)}$ is the intensity at the output port with constructive (destructive) interference. The sorting fidelities are $(97.5\pm 0.1)\%$ and $(96.4\pm 0.1)\%$ for $|l=1\rangle+|l=-1\rangle$ and $|l=2\rangle+|l=-2\rangle$, respectively.

\begin{figure*}[!hbt]
\centering
\resizebox{15.2cm}{2.8cm}{\includegraphics{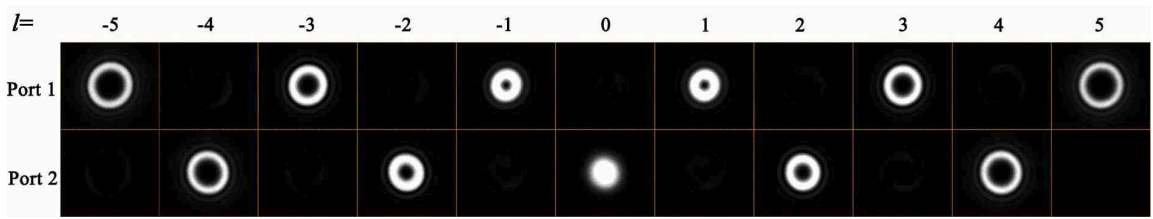}}
\caption{The images of the OAM modes output from Port 1 and Port 2. Left to right: $|l=-5\rangle$ to $|l=5\rangle$.}
\label{fig4}
\end{figure*}

\begin{figure}[!hbt]
\centering
\resizebox{7.5cm}{4.5cm}{\includegraphics{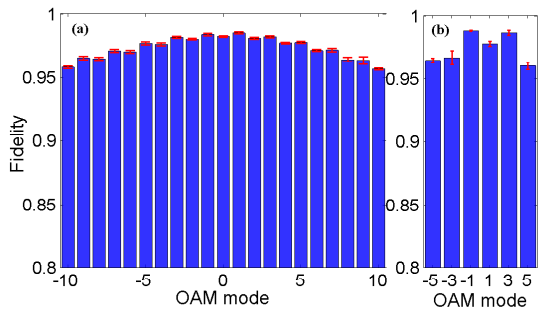}}
\caption{The sorting fidelities of the OAM filter for different modes with (a) PMM1 only and (b) PMM2 cascaded after PMM1, respectively. The error bar is set as the standard deviation.}
\label{fig5}
\end{figure}

The output polarization of the even-order OAM modes is orthogonal to that of the odd-order ones when  $\alpha_1=\alpha_2=\pi/4$. Then, the PMM becomes a 2-dimensional controlled-NOT gate. By cascading the detection module afterward, the controlled-NOT gate can be used as an OAM filter, which filters the odd-order modes from the even-order ones. The images of eleven different OAM modes output from Port 1 and 2 are shown in Fig. \ref{fig4}. Figure \ref{fig5}(a) gives the sorting fidelities of PMM1 for different modes and error bars are set as one standard deviation. The sorting fidelity is $\mathcal{F}=(98.5\pm 0.1)\%$ when $l=1$ and decreases slightly as $|l|$ increases due to experimental errors. That is why $\mathcal{F}=(95.7\pm 0.1)$ when $l=10$. We also implement a two-stage PMM structure by cascading a second PMM (PMM2, as shown in Fig. \ref{fig2}(b)) after Port 1 in Fig. \ref{fig2}(a). PMM2 is utilized to filter $4n+1$-order OAM modes from $4n+3$-order ones. Thus, according to Ref. \cite{Leach2002}, the rotation angles of the DPs (DP3 and DP4) of PMM2 are set as $\alpha_3=\alpha_4=\pi/8$, the rotation angles of the HWPs on both sides of DP3 (DP4) are set as $\pi/16$ ($-\pi/16$) and the quarter wave-plate (QWP) with  zero rotation angle before PMM2 introduces a $\pi/2$ phase shift to $|V\rangle$. The corresponding sorting fidelities are shown in Fig. \ref{fig5}(b). As the experimental errors are amplified in the cascading structure, the sorting fidelity of PMM2 becomes $\mathcal{F}=(96.0\pm 0.3)\%$ when $l=5$. 

\begin{figure}[!hbt]
\centering
\resizebox{5.71cm}{4cm}{\includegraphics{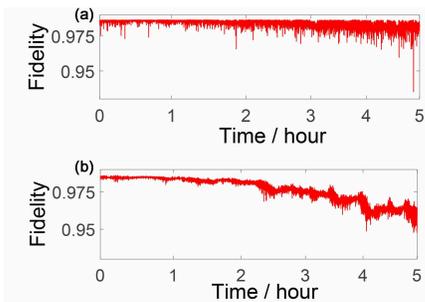}}
\caption{The sorting fidelities of the OAM filter for $|l=1\rangle$ with (a) PMM1 only and (b) PMM2 cascaded after PMM1, respectively.}
\label{fig6}
\end{figure}

Figure \ref{fig6} gives the free-running sorting fidelities of PMM1 (Fig. \ref{fig6}(a)) and the 2-stage cascading structure (Fig. \ref{fig6}(b)) for $l=1$. The sorting fidelity of PMM1 is still as high as 98.2\% even after 5 hours' free running. The sorting fidelity of the 2-stage cascading structure decreases faster but is still about 95.9\% after 5 hours' free running. The sorting fidelities of both Figs. \ref{fig6}(a) and \ref{fig6}(b) are nearly invariant within the first hour. Thus, the PMM is suitable for cascading structures and is stable for free running.

The major experimental error comes from the rotators utilized in the experiment. The rotation precisions of rotators of all DPs and wave plates are 2 degree and the practical output polarization of the PMM is $|H\rangle+e^{-i4l(\alpha+\delta)}|V\rangle$, where $\delta$ is the rotating error of the DP. Then, we can obtain that $\mathcal{F}_{l,\delta}=(1+cos(4l\delta))/2$. As $\mathcal{F}=(95.7\pm 0.1)\%$ for $l=10$, it can be obtained that $\delta\simeq 0.6$ degree. The experimental error is consistent with the rotation precision. The generation purity of the SLM and the rotation errors of different modes are superposed, thus the sorting fidelities of the superposition states are slightly lower. The sorting fidelity can be improved further by utilizing rotators with higher precision. 

It should be noted that though the polarization output from of a normal PBS Sagnac interferometer with a DP can be compensated by generic wave plates \cite{Slussarenko2010,D'Ambrosio2012}, the polarization state output depends not only on $l$ and $\alpha$ but also on the specific parameters of the DP \cite{Wang2017}. Thus, when the dimensionality is larger than two, the compensation can not be satisfied simultaneously for  all OAM modes to perform the high-dimensional C-phase gate even by cascading a second DP after the Sagnac interferometer. The polarization output from the PMM proposed here is independent on the specific parameters of the DP and is universal to all OAM modes, which means the PMM is a feasible tool for practical high-dimensional quantum and classical optical communication with OAM. Additionally, the practical single photon source with a certain spectrum line width within 10 $nm$ will not lead to a significant error \cite{Yao2012}, as only the wave plates in the PMM are sensitive to the spectrum line width.

In conclusion, we have implemented a high-dimensional controlled-phase gate for hybrid OAM photon states by a compact PMM, which can compensate the mode-dependent global phase automatically and preserve the phase in the spin-OAM hybrid superposition state. The free running PMM is suitable for high-dimensional quantum information processing and classical optical communication.

\section*{Funding Information}

This work has been supported by the National Natural Science Foundation of China (Grant Nos. 61675189, 61627820, 61622506, 61475148, 61575183), the National Key Research And Development Program of China (Grant Nos.2016YFA0302600, 2016YFA0301702), the "Strategic Priority Research Program(B)" of the Chinese Academy of Sciences (Grant No. XDB01030100).

\end{document}